\documentclass[prl,twocolumn,superscriptaddress,floatfix]{revtex4}
\usepackage[utf8]{inputenc}
\pdfoutput=1
\usepackage{graphicx}
\usepackage{float}
\usepackage{amsmath,amssymb,amstext,dsfont,tikz,graphicx,color,soul,physics,xfrac}

\providecommand{\ignore}[1]{}
\newcommand{\Z}{\mathcal{Z}}
\newcommand{\T}{\mathcal{T}}
\newcommand{\A}{\mathcal{A}}
\newcommand{\B}{\mathcal{B}}
\newcommand{\C}{\mathcal{C}}
\newcommand{\id}{\mathds{1}}
\newcommand{\Ca}{\mathsf{C}_a}
\newcommand{\Sa}{\mathsf{S}_a}
\newcommand{\Ta}{\mathsf{T}_a}
\newcommand{\Cb}{\mathsf{C}_b}
\newcommand{\Sb}{\mathsf{S}_b}
\newcommand{\Tb}{\mathsf{T}_b}

\newcommand{\tc}{\text{4DTC}}

\begin{document}

\title{Universality Classes of Stabilizer Code Hamiltonians}
\author{Zack Weinstein}
\affiliation{Department of Physics, Washington University, St.\ Louis, MO 63130, USA}

\author{Gerardo Ortiz}
\affiliation{Department of Physics, Indiana University, Bloomington, IN 47405, USA}

\author{Zohar Nussinov}
\affiliation{Department of Physics, Washington University, St.\ Louis, MO 63130, USA}

\date{\today}

\begin{abstract}
Stabilizer code quantum Hamiltonians have been introduced with the intention of physically realizing a quantum memory because of their resilience to decoherence. In order to analyze their finite temperature thermodynamics, we show how to generically solve their partition function using duality techniques. By unveiling each model’s universality class and effective dimension, insights may be gained on their finite temperature dynamics and robustness. Our technique is demonstrated in particular on the 4D toric code and Haah’s code -- we find that the former falls into the 4D Ising universality class, whereas Haah’s code exhibits dimensional reduction and falls into the 1D Ising universality class.
\end{abstract}

\maketitle

The stabilizer formalism is a powerful mathematical framework for designing quantum error correcting codes \cite{shor_scheme_1995,calderbank_good_1996,gottesman_stabilizer_1997,gottesman_theory_1998, bacon_operator_2006, cafaro_approximate_2014}. Kitaev proposed to turn a stabilizer code into an interacting many-body system by associating a coding space to the ground state subspace of a stabilizer code Hamiltonian, a linear combination of elements of the stabilizer group
\cite{kitaev_fault-tolerant_2003,dennis_topological_2002}. Stabilizer code Hamiltonians display a gapped spectrum with a topologically quantum ordered ground manifold and where errors, typified by finite energy excitations, become energetically unfavorable at zero temperature. These systems are natural candidates for physical realization of a robust quantum memory, a q-RAM or q-Hard Drive, because of their inherent resilience to decoherence. Since Kitaev's original proposal, several stabilizer code models have been advanced in various spatial dimensions $D$, including the most recent {\it fracton} models \cite{vijay_fracton_2016, shirley_foliated_2018, prem_glassy_2017, nandkishore_fractons_2018, haah_local_2011, haah_lattice_2013}.

It was emphasized long ago that the effect of temperature on these memories cannot be ignored \cite{nussinov_autocorrelations_2008, alicki_statistical_2007, bravyi_no-go_2009}. Finite temperature decoherence times may be affected by effective {\it dimensional reduction}  \cite{nussinov_symmetry_2009, nussinov_effective_2012, nussinov_autocorrelations_2008}. Specifically, the spectral degeneracy of stabilizer code models is associated with symmetries that may involve a macroscopic fraction of degrees of freedom, the so-called $d$-dimensional gauge-like symmetries \cite{batista_generalized_2005, nussinov_sufficient_2009}, $0 \le d \le D$, later on dubbed ``subsytem'' symmetries \cite{gaiotto_generalized_2015, lake_higher-form_2018, devakul_classification_2018, williamson_spurious_2019}. Duality transformations \cite{kramers_statistics_1941,wegner_duality_1971,cobanera_bond-algebraic_2011} may unveil the lower dimensional {\it classical} theory isomorphic to the stabilizer code model. Such dual theories exhibit non-analyticities (and critical exponents associated with continuous transitions \cite{Nishimori-Ortiz-2011}) of identical character, and therefore belong to the same universality class. Understanding the universality classes and dynamics of stabilizer models may aid in the design of robust quantum memories. 

The primary goal of this paper is to show how 
duality techniques can be utilized to exactly determine the partition function of stabilizer code Hamiltonians. In particular, we demonstrate how the stabilizer algebra encodes any non-analyticities (or lack thereof) in the thermodynamic free energy of the corresponding stabilizer Hamiltonian, via the scaling of constraints on the stabilizer algebra with system size. The Abelian nature of the stabilizer group allows for a particularly simple analysis: while the studied models are in principle constructed using a large number of entangled quantum spins, the resulting algebra will be shown to factor into independent Ising algebras. Consequently, the partition function of \textit{any} CSS stabilizer code Hamiltonian \cite{kitaev_classical_2002} may be easily analyzed using various duality techniques. The effective dimensionality of the resulting classical models vary depending on the constraints. We find that $D=2$ or $3$ dimensional stabilizer models are often dual to classical Ising chains, implying the absence of phase transitions in many stabilizer models  (see Table \ref{tab:1}). 

In this paper, we analyze the 4D toric code \cite{dennis_topological_2002} and Haah's 3D cubic code \cite{haah_local_2011,haah_lattice_2013,bravyi_analytic_2013}. These models represent two extremes of the dimensional reduction paradigm: we will show that the 4D toric code (4DTC) features no dimensional reduction and belongs to the 4D Ising universality class. By contrast, typical odd lattice size renditions of Haah's 3D cubic code lie in the 1D Ising universality class. The 4DTC therefore exhibits a finite temperature phase transition with critical exponents given exactly by those of mean field theory, while Haah's cubic code may be unstable to thermal fluctuations (a phenomenon known as {\it thermal fragility} \cite{nussinov_autocorrelations_2008}) and exhibit no finite temperature transitions.

\begin{table*}[t]
  \centering
  \begin{tabular}{|l|l|l|l|l|}
\hline
Model                   & $D$ & $d$ & Dual Model & Universality Class   \\ \hline
2D toric code \cite{kitaev_fault-tolerant_2003, nussinov_autocorrelations_2008} & 2 & 1 & Two decoupled 1D Ising chains & 1D Ising \\ \hline
2D honeycomb toric code \cite{liang_anyon_2011,nussinov_effective_2012} & 2 & 1 & Two decoupled 1D Ising chains & 1D Ising \\ \hline
Color codes \cite{bombin_topological_2006,nussinov_effective_2012} & 2 & 1 & Two decoupled 1D Ising chains & 1D Ising \\ \hline
3D toric code \cite{castelnovo_topological_2008,nussinov_autocorrelations_2008} & 3 & 0, 1 & Decoupled 1D Ising and 3D Ising models & 3D Ising  \\ \hline
X-Cube* \cite{vijay_fracton_2016,weinstein_absence_2018} & 3 & 1, 2 & Decoupled $L$ 1D Ising and $L-1$ 1D Ising-gauge & 1D Ising \\ \hline
Haah's code** \cite{haah_local_2011,haah_lattice_2013,bravyi_analytic_2013} & 3 & 2 & Two decoupled 1D Ising chains & 1D Ising \\ \hline
4D toric code \cite{dennis_topological_2002,alicki_thermal_2008} & 4 & 2 & Two decoupled 4D Ising models & 4D Ising \\ \hline
Chamon's XXYYZZ \cite{chamon_quantum_2005,cobanera_bond-algebraic_2011,nussinov_effective_2012} & 3 & 1 & Four decoupled 1D Ising chains & 1D Ising \\ \hline
\end{tabular}
  \caption{Universality classes of stabilizer code Hamiltonians. $D$ is the spatial dimension of the lattice model. $d$ is the dimension of the gauge-like symmetries. Dualities are defined as equivalence relations between partition functions: the 3DTC, for example, has a partition function proportional to the product of a 1D Ising and a 3D Ising partition function. While Chamon's XXYYZZ model is not a stabilizer code, it can also be shown by duality to exhibit dimensional reduction. Additionally, while all listed models above are constructed using Pauli operators, very similar results may be obtained for non-Pauli models, such as those with $\mathbb{Z}_p$ clock operators or $U(1)$ operators. *: While the X-Cube model's universality class does not depend on any choice of boundary conditions, the particular duality chosen holds for the case of cylindrical boundary conditions. **: The duality given below for Haah's code holds explicitly for those values of $L$ for which the ground state degeneracy (GSD) is 4.}
  \label{tab:1}
\end{table*}

\textit{Methodology -- } We will investigate thermal properties of the above two stabilizer Hamiltonians by identifying their classical Ising duals. The models are defined on $D$-dimensional lattices $\Lambda = \mathbb{Z}_L^D$ of length $L$ in each direction with vertices $v = (x,y,z,\ldots) \in \Lambda$. The lattices $\Lambda$ are endowed with periodic boundary conditions, although any local finite temperature properties should not depend on the boundary conditions in the thermodynamic limit. We associate with the lattice $N$ qubits, each with a local Hilbert space $\mathcal{H}_n = \mathbb{C}^2$; the global Hilbert space $\mathcal{H} = \otimes_n \mathcal{H}_n$ is of complex dimension $2^N$. Each qubit belongs to a unit $k$-cell of the lattice: $k=0,1,$ and $2$ represent qubits on the vertices, links, and plaquettes of the lattice respectively. Our arguments are easily generalized to $\mathsf{p}$-qudits and $U(1)$ models \cite{anwar_fast_2014,weinstein_absence_2018}. 

Next, we define the operators $A_r$ and $B_s$ as:
\begin{equation}
\label{eq:ABdefn}
\begin{split}
A_r \equiv \prod_{n \in N_r} \sigma^x_n , \quad 1 \leq r \leq R , \\
B_s \equiv \prod_{n \in N_s} \sigma^z_n , \quad 1 \leq s \leq S,
\end{split}
\end{equation}
where $N_r$ and $N_s$ are indexing sets used to generate $R$ operators $A_r$ and $S$ operators $B_s$, respectively. We further require that each $A_r$ and $B_s$ commute.
% \begin{equation}
% \label{eq:ABcommute}
% [ A_r, B_s ] = 0 \ \forall r,s.
% \end{equation}
The Hamiltonian for this generic stabilizer model reads
\begin{equation}
\label{eq:Hdefn}
H = -a\sum_{r=1}^R A_r - b\sum_{s=1}^S B_s ,
\end{equation}
with coupling constants $a,b>0$. All operators in (\ref{eq:Hdefn}) commute and square to the identity $\id$ on $\mathcal{H}$. The partition function is then given by the following high-temperature ($\beta=1/(k_B T)$) series expansion:
\begin{equation}
\label{eq:Zdefn}
\begin{split}
\Z &= \Tr e^{-\beta H} = \Tr \left[ \prod_{r=1}^R (\id \Ca + A_r \Sa) \prod_{s=1}^S (\id \Cb + B_s \Sb) \right] \\
&= 2^N \Ca^R \Cb^S \T_a \T_b.
\end{split}
\end{equation}
Here, $\Ca \equiv \cosh(\beta a)$ and $\Sa \equiv \sinh(\beta a)$, with $\Cb$ and $\Sb$ similarly defined. In the above, $\T_a$ (and analogously $\T_b$) are given by
\begin{equation}
\label{eq:Tdefn}
\prod_{r=1}^R [\id + A_r \Ta] = \T_a \id + {\sf t.t.} \ , 
\mbox{with } \T_a = \sum_{P \in \A} \Ta^{|P|},
\end{equation}
with ``{\sf t.t.}" denoting traceless terms and $\Ta \equiv \tanh(\beta a)$ (and $\Tb \equiv \tanh(\beta b)$). $P \in \A$ ($\B$) denotes operators $A_r$ ($B_s$) multiplying to $\id$,
\begin{equation}
\label{eq:Aproduct}
\prod_{\ell \in P} A_{\ell} = \id \quad \forall P \in \A .
\end{equation}
Each $P$ corresponds to a \textit{constraint} on the stabilizer algebra. The only terms contributing to the trace in (\ref{eq:Zdefn}) are those proportional to the identity ($2^N = \Tr[\id]$). The traceless terms ({\sf t.t.}) in (\ref{eq:Tdefn}) and those corresponding to $\T_b$ cannot combine to yield the identity -- by construction in (\ref{eq:ABdefn}), there are no nontrivial constraints between $A_r$ and $B_s$ operators. We have thus reduced the problem of solving each model's partition function to identifying which and how many constraints exist among $A_r$ or $B_s$ operators separately. From the partition function, we may then compute the thermodynamic free energy density, 
\begin{equation}
    f(\beta) = \lim_{L \rightarrow \infty} - \frac{1}{\beta L^D} \log \Z .
\end{equation}

This means of describing the thermodynamics of spin models is particularly effective for stabilizer Hamiltonians. The algebra of a stabilizer Hamiltonian has three important properties: (i) each element of the stabilizer commutes with one another; (ii) each element of the stabilizer is usually composed of either entirely $\sigma^x$ or $\sigma^z$ operators (these stabilizer codes are known as CSS codes \cite{kitaev_classical_2002}, and most stabilizer code Hamiltonians fall into this category); (iii) each element has eigenvalues $\pm 1$. This implies that the stabilizer algebras that we investigate factor into two classical Ising algebras. As a result, these stabilizer Hamiltonians are dual \cite{kramers_statistics_1941, wegner_duality_1971, cobanera_bond-algebraic_2011} to classical Ising-like Hamiltonians using bond-algebraic dualities \cite{cobanera_bond-algebraic_2011}. %The same strategy can be easily implemented on $\mathsf{p}$-qudits, in which Pauli operators are replaced by elements of the Weyl algebra \cite{weinstein_absence_2018}.

\textit{4D toric code -- } As befits its name, the 4DTC \cite{dennis_topological_2002, duivenvoorden_renormalization_2019} is defined on a $D=4$ dimensional lattice. Qubits are associated with all ($6L^{4}$) plaquettes $p$. For each link $\ell$, the operator $A_{\ell}$ is defined by 
\begin{equation}
\label{eq:4dtcA}
A_{\ell} \equiv \prod_{\ell \in \partial p} \sigma^x_p ,
\end{equation}
where the above product is over the six plaquettes $p$ whose boundary $\partial p$ contains the link $\ell$. The operator $B_c$ is defined for each three-dimensional cube as
\begin{equation}
\label{eq:4dtcB}
B_c \equiv \prod_{p \in \partial c} \sigma^z_p.
\end{equation}
The above product is over the six plaquettes contained in the cube $c$'s boundary. The Hamiltonian $H_{\text{4DTC}}$ and partition function $\Z_{\text{4DTC}}$ are as defined in (\ref{eq:Hdefn}) and (\ref{eq:Zdefn}) respectively, and it is trivially verified that each $A_{\ell}$ and $B_c$ commute.

We now show that the 4DTC is dual to two copies of the 4D nearest neighbor Ising (4DI) model defined by 
\begin{equation}
\label{eq:HIsing}
H_{\text{4DI}} = -J \sum_{\langle v, v' \rangle} s^{\;}_v s_{v'},
\end{equation}
in the sense that the thermodynamic free energy density can be trivially written in terms of the 4D Ising model's free energy. In (\ref{eq:HIsing}), $s_v$ is a classical spin variable at each vertex $v \in \Lambda$ and the sum is over all nearest neighbor pairs $v, v'$ in $\Lambda$. We express the partition function of this model via a \textit{low} temperature series expansion. Starting from the ground state of $s_v = +1$ for all $v$, we consider excited states and expand in the number of higher energy ``broken bonds"; a ``broken bond" corresponds to $s^{\;}_v s_{v'} = -1$ for a nearest neighbor pair $v, v'$. The partition function is then given by
\begin{equation}
\label{eq:ZIsing}
\Z_{\text{4DI}} = 2 e^{4L^4 \beta J} \sum_{\C \subset \Lambda} e^{-2\beta J \Delta_{\C}}.
\end{equation}
Each $\C$ represents a set of flipped spins from the chosen ground state. We demand $| \C | \leq L^4 / 2$, noting that each configuration $\C$ has a global spin-flip ``symmetry partner" $\Lambda \setminus \C$ with the same bond structure as $\C$. The ground state energy is $-4L^4 J$, and $\Delta_{\C}$ is the number of ``broken bonds" in configuration $\C$ with $2J$ being the energy penalty for ``breaking a bond".  

\begin{figure}
\begin{center}
\includegraphics[scale=.295]{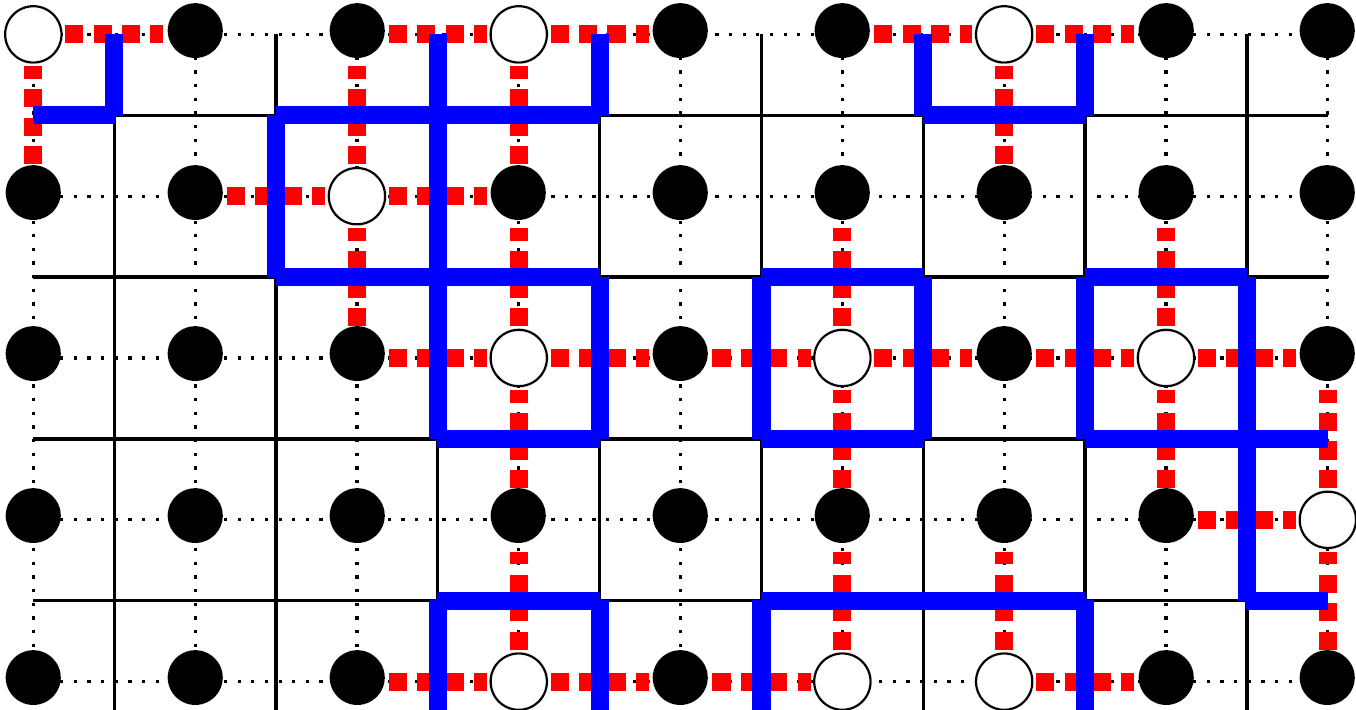}
\end{center}
\caption{A 2D cross section of a 4D lattice, with classical Ising spins at each site. Each red dotted link corresponds to a broken bond, and each blue loop is the 1D cross-section of a 3D hypersurface domain wall. In the 4D lattice, the product of $A_{\ell}$ over all red links and the product of $B_{c}$ over all blue cubes correspond to two independent constraints.}
\label{fig:Bduality}
\end{figure}

We begin investigating $\T_a$ by noting the constraint
\begin{equation}
\label{eq:4DTCAconstraint}
\prod_{v \in \partial \ell} A_{\ell} = \id.
\end{equation}
The above product is over the eight links containing the vertex $v$. This can be verified by noting that there are twenty four plaquettes adjacent to $v$ (four for each $\mu$-$\nu$ plane), and each plaquette is included by exactly two links in (\ref{eq:4DTCAconstraint}). Higher order constraints may be found by taking products of (\ref{eq:4DTCAconstraint}) for some subset $\C$ of vertices, and eliminating any $A_{\ell}$ included in the product twice. Note that each $A_{\ell}$ for a given link $\ell = \{ v, v' \}$ will be included in such a product if and only if $v \in \C$ or $v' \in \C$, so that $\C$ and $\Lambda \setminus \C$ yield the same constraint (see Fig. \ref{fig:Bduality}).

This set of constraints suggests the following duality: for each spin flip configuration $\C$ in the Ising low temperature expansion (\ref{eq:ZIsing}), we obtain a unique identity product in the 4DTC high temperature expansion. Moreover, each $A_{\ell}$ operator included in such a constraint must correspond exactly to a bad bond in $\C$. This shows that the series (\ref{eq:ZIsing}) is entirely contained within (\ref{eq:Tdefn}), with a global prefactor and the replacement $e^{-2\beta J} \rightarrow \Ta$: 
\begin{equation}
\label{eq:4DTCAduality}
\T_a = \frac{1}{2} \Ta^{2L^4} \Z_{\text{4DI}} \left( \frac{1}{2J} \log \frac{1}{\Ta} \right) + \mathcal{O}(\Ta^{L^3}),
\end{equation}
where the terms of order $\Ta^{L^3}$ or higher, $\mathcal{O}(\Ta^{L^3})$, arise due to the topology of the lattice, and are negligible to the thermodynamic free energy.

Turning to $\T_b$, we similarly note the constraint
\begin{equation}
\label{eq:4DTCBconstraint}
\prod_{c \in \partial h} B_c = \id ,
\end{equation}
where the product is over the eight cubes contained in a minimal 4D hypercube $h$. By multiplying such constraints, we can generalize (\ref{eq:4DTCBconstraint}) to include products over any closed three dimensional hypersurface; the number of $B_c$ operators in such products is the three dimensional hypersurface area (Fig. \ref{fig:Bduality}). This set of constraints also suggests a duality to (\ref{eq:ZIsing}) via another lens. Here, instead of placing Ising spins at each vertex of the lattice, we imagine placing spins at the \textit{center} of each hypercube $h$, creating another lattice $\Lambda'$ a half-spacing off from $\Lambda$. Whereas the $A_{\ell}$ in  (\ref{eq:4DTCAconstraint}) stem from broken bonds in the Ising model on the same lattice, each $B_c$ operator in this duality represents a \textit{domain wall} separating different spin orientations in an Ising model on $\Lambda'$. The hypersurface area of this domain wall is equal to the number of broken bonds in the Ising configuration. From this, we see that the same duality as (\ref{eq:4DTCAduality}) holds for $\T_b$, with $\Ta$ replaced with $\Tb$. Indeed, once $\Z_{\text{4DTC}}$ has been factored as in (\ref{eq:Zdefn}), this duality is known as a Wegner duality \cite{wegner_duality_1971}, or more particularly as a ``lattice gerbe theory" duality \cite{johnston_z_2_2014}.

One might reasonably worry that the above discussion is too cavalier: although it's clear that each pair of spin flip configurations $\C$ and $\Lambda \setminus \C$ generates a unique constraint via (\ref{eq:4DTCAconstraint}), how do we know that all terms of (\ref{eq:Tdefn}) below order $\Ta^{L^3}$ can be found this way? Additionally, how do we know that the subextensive contributions to $\Z_{\tc}$ are negligible in the free energy's thermodynamic limit? These questions, and analogous ones for $\T_b$, are addressed with a careful proof of the duality (\ref{eq:4DTCAduality}) in the supplemental material.

\begin{figure}
\begin{center}
\includegraphics[scale=1]{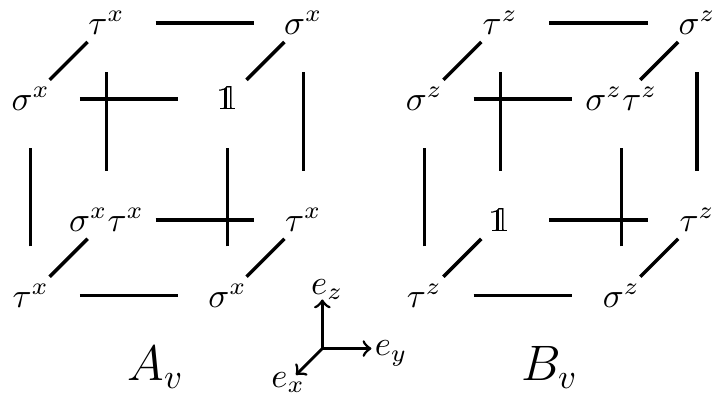}
%\begin{tikzpicture}
%\foreach \x in {0,2} {
%	\foreach \y in {0,2} {
%		\draw[line width = 1.0pt, black] (\x,\y,0.5) -- (\x,\y,1.5);
%		\draw[line width = 1.0pt, black] (\x,0.5,\y) -- (\x,1.5,\y);
%		\draw[line width = 1.0pt, black] (0.5,\x,\y) -- (1.5,\x,\y);
%		\draw[line width = 1.0pt, black] (\x+4,\y,0.5) -- (\x+4,\y,1.5);
%		\draw[line width = 1.0pt, black] (\x+4,0.5,\y) -- (\x+4,1.5,\y);
%		\draw[line width = 1.0pt, black] (4.5,\x,\y) -- (5.5,\x,\y);	
%	}
%}
%\node at (0,0,0) {$\sigma^x \tau^x$};
%\node at (2,0,0) {$\tau^x$};
%\node at (0,2,0) {$\tau^x$};
%\node at (0,0,2) {$\tau^x$};
%\node at (2,2,0) {$\sigma^x$};
%\node at (2,0,2) {$\sigma^x$};
%\node at (0,2,2) {$\sigma^x$};
%\node at (2,2,2) {$\mathds{1}$};
%
%\node at (4,0,0) {$\mathds{1}$};
%\node at (6,0,0) {$\tau^z$};
%\node at (4,2,0) {$\tau^z$};
%\node at (4,0,2) {$\tau^z$};
%\node at (6,2,0) {$\sigma^z$};
%\node at (6,0,2) {$\sigma^z$};
%\node at (4,2,2) {$\sigma^z$};
%\node at (6,2,2) {$\sigma^z \tau^z$};
%
%\node at (0.5,-1.5,0) {\Large $A_v$};
%\node at (4.5,-1.5,0) {\Large $B_v$};
%
%\draw[->, line width = 0.8pt] (2,-1.5,-0.5) -- (2,-1,-0.5);
%\draw[->, line width = 0.8pt] (2,-1.5,-0.5) -- (2.5,-1.5,-0.5);
%\draw[->, line width = 0.8pt] (2,-1.5,-0.5) -- (2,-1.5,0.2);
%\node at (2,-1.5,0.4) {$e_x$};
%\node at (2.7,-1.55,-0.5) {$e_y$};
%\node at (2.05,-0.9,-0.5) {$e_z$};

%\end{tikzpicture}
\end{center}
\caption{Haah's code: the two operators of Eqs. (\ref{eq:HaahA}), (\ref{eq:HaahB}).}
\label{fig:haah}
\end{figure}

\textit{Haah's code -- } Haah's code is defined as follows \cite{haah_local_2011, nandkishore_fractons_2018}: Let $\Lambda$ be a $D=3$ lattice, where we associate \textit{two} qubits with each vertex $v \in \Lambda$. Letting $\sigma^{\mu}_v$ and $\tau^{\mu}_v$ label the first and second qubits respectively at each vertex, the operator $A_v$ is then defined as in Fig. \ref{fig:haah}:
\begin{equation}
\label{eq:HaahA}
\begin{split}
A_v \equiv & \sigma^x_v \tau^x_v \tau^x_{v + e_x} \tau^x_{v + e_y} \tau^x_{v + e_z} \\ 
{} & \sigma^x_{v + e_x + e_y} \sigma^x_{v + e_x + e_z} \sigma^x_{v + e_y + e_z} .
\end{split}
\end{equation}
The operator $B_v$ is similarly defined as in Fig. \ref{fig:haah}:
\begin{equation}
\label{eq:HaahB}
\begin{split}
B_v \equiv & \tau^z_{v+e_x} \tau^z_{v+e_y} \tau^z_{v+e_z} \sigma^z_{v+e_x+e_y} \\
{} & \sigma^z_{v+e_x+e_z}\sigma^z_{v+e_y+e_z}\sigma^z_{v+e_x+e_y+e_z}\tau^z_{v+e_x+e_y+e_z} .
\end{split}
\end{equation}
As usual, $[A_v , B_{v'}] = 0$ for any two sites $v$ and $v'$, and the model's Hamiltonian $H_{\text{Haah}}$ and partition function $\Z_{\text{Haah}}$ are defined as in (\ref{eq:Hdefn}) and (\ref{eq:Zdefn}), respectively. Note that the operators $A_v$ and $B_v$ are simply reflections of one another, so their constraints will be identical.

Haah's code features an intricate ground state degeneracy (GSD) \cite{haah_lattice_2013}: unless $L$ is a multiple of $4^p - 1$ for $p \geq 2$, $\text{GSD}=4$ for odd $L$ \cite{haah_local_2011}. We will restrict our attention to these models, as they are the most pertinent to quantum error correction: it has been argued \cite{haah_lattice_2013, bravyi_analytic_2013} that, for these values of $L$, the model demonstrates long memory timescales at low temperatures.

While the nature of constraints in $H_{\text{Haah}}$ varies wildly for different values of $L$, the number of constraints (including the trivial empty product) is always equal to $\log_{2} \text{GSD}$  \cite{haah_lattice_2013}. Thus, when $\text{GSD}= 4$, the two independent constraints are those present for all $L$,
\begin{equation}
\label{eq:HaahConstraints}
\prod_{v \in \Lambda} A_v = \id , \quad \prod_{v \in \Lambda} B_v = \id .
\end{equation}
These relations can easily be verified by observing that the product of each of the eight corners of the cubic operators yields the identity. The partition function is then:
\begin{equation}
\label{eq:Zhaah}
\Z_{\text{Haah}} = 2^{2L^3} \left( \Ca^{L^3} + \Sa^{L^3} \right) \left( \Cb^{L^3} + \Sb^{L^3} \right) .
\end{equation}
Alternatively, let $s_i$ and $ t_i$ be classical Ising spins ($1 \leq i \leq L^3$). Then, within the bond-algebraic framework of dualities as isomorphisms
\cite{cobanera_bond-algebraic_2011}, the mapping
\begin{equation}
\label{eq:HaahDuality}
A_v \rightarrow s_i s_{i+1} , \quad B_v \rightarrow t_i t_{i+1} , \quad 1 \leq i \leq L^3
\end{equation}
with $L^3+1 \equiv 1$ similarly implies the duality of Haah's code to two periodic Ising chains. This duality suggests that the finite temperature dynamics of Haah's code are identical to those of finite temperature classical Ising chains, and may thus be unstable to thermal fluctuations (i.e., exhibit ``thermal fragility'' \cite{nussinov_autocorrelations_2008}).

We caution that the duality (\ref{eq:HaahDuality}) only holds exactly for the case $\text{GSD}= 4$: although the thermodynamic limit can be taken along an infinite sequence of $L$ satisfying this condition \cite{bravyi_analytic_2013}, each variation of constraints at different $L$ requires a new bond-algebraic duality. Nevertheless, the number of constraints never exceeds $2^{\mathcal{O}(L)}$ for any system size, and the free energy remains analytic in any thermodynamic limit as a result \cite{haah_lattice_2013}. This suggests that Haah's code may remain in the 1D Ising universality class along any thermodynamic limit.

\textit{Conclusions -- } 
We showed how to generically analyze the partition function of a CSS stabilizer code Hamiltonian using duality techniques. We illustrated our strategy on the 4DTC and Haah's cubic code, two quintessential stabilizer codes. Our results support the generally held belief that the 4DTC exhibits self-correcting properties at sufficiently low temperatures. While several works suggested that Haah's code may be partially self-correcting at finite temperatures \cite{haah_local_2011,haah_lattice_2013,bravyi_analytic_2013,brown_quantum_2016}, our results instead suggest that Haah's code may suffer the same thermal fragility as Ising chains. While the dimensional reduction implied by generalized Elitzur's theorem \cite{batista_generalized_2005, nussinov_effective_2012} bounds correlation functions on $d$-dimensional subsystems, this does not imply that the thermodynamics is that of canonical $d$-dimensional systems (c.f., (a) the 90$^{\circ}$ square lattice compass model \cite{nussinov_compass_2015}, a system with $d=1$ symmetries and 2D Ising behavior or (b) the ``XXZ honeycomb model''\cite{PhysRevB.86.085415}, a compass model with similar $d=1$ symmetries, that is dual to the 2D quantum Ising lattice gauge theory).

The 2D Ising model can serve as a self-correcting classical memory below its critical temperature, in the sense that the information stored is robust to magnetic or thermal fluctuations \cite{day_ising_2012, poulin_self-correction_2019}. By contrast, the 1D Ising model suffers a finite memory timescale independent of system size at all nonzero temperatures due to the absence of an ordered phase \cite{song_high_2009, day_ising_2012}. Similarly, it is commonly accepted that the 2DTC suffers from relaxation times independent of system size, while the 4DTC is believed to function as a robust quantum memory below a critical temperature \cite{nussinov_autocorrelations_2008,alicki_statistical_2007,alicki_thermal_2008,alicki_thermalization_2009}. The relationship between these classical and quantum memories can be understood through duality: bond algebraic dualities suggest that the dynamics of the 2DTC on the torus are identical to that of two 1D Ising chains. In the supplemental material, we discuss how the topological degeneracy commonly associated with these models appears as a global prefactor in the dual partition function. A common concern regarding the use of dualities for analyzing dynamics is that local heat bath perturbations become generally nonlocal in the dual model. In the supplemental material, we show that a local coupling to a heat bath in the 2DTC can induce a local coupling in the dual Ising model as well.

This analogy highlights the utility of the duality techniques developed in this paper: by determining a stabilizer code Hamiltonian’s classical dual and corresponding universality class, one obtains all information regarding the model's critical phenomena without performing detailed numerical analyses (see Table \ref{tab:1}). Using the techniques explicitly demonstrated here and in previous work, we conjecture that all sufficiently generic stabilizer models -- CSS and beyond -- can be analyzed for thermodynamic and, in some cases, dynamical behaviors. A finite temperature phase transition and corresponding stable phase may be a crucial ingredient \cite{bacon_operator_2006, nussinov_autocorrelations_2008} for large autocorrelation times and robust quantum memories.

\textit{Acknowledgements -- } This work was partially supported by the National Science Foundation under grant NSF 1411229 and by the U.S. Department of Energy under grant DE-SC0020343.

\end{document}

\title{Universality Classes of Stabilizer Code Hamiltonians: Supplemental Material}
\author{Zack Weinstein}
\affiliation{Department of Physics, Washington University, St.\ Louis, MO 63130, USA}

\author{Gerardo Ortiz}
\affiliation{Department of Physics, Indiana University, Bloomington, IN 47405, USA}

\author{Zohar Nussinov}
\affiliation{Department of Physics, Washington University, St.\ Louis, MO 63130, USA}

\date{\today}

\begin{abstract}

\end{abstract}

\maketitle

\section{Gauge-like Symmetries}
We briefly review the $d$-dimensional gauge-like symmetries of the 4DTC and Haah's Code. These symmetries may act as logical operators on the ground state manifold, manipulating the encoded logical qubits. Two easily constructed ground states can be written in the language of the main paper:
\begin{equation}
    \label{eq:psiphidefn}
    \begin{split}
    \ket{\psi_0} = \prod_{s=1}^S \frac{1}{2} (\id + B_s) \ket{+x} , \\ \ket{\phi_0} = \prod_{r=1}^R \frac{1}{2} (\id + A_r) \ket{+z} ,
    \end{split}
\end{equation}
where $\ket{+x}$ and $\ket{+z}$ respectively denote the simultaneous $+1$ eigenstate of each $\sigma^x_n$ and $\sigma^z_n$.

\subsection{4D Toric Code}
The 4DTC displays two $d=2$ dimensional gauge-like symmetries \cite{duivenvoorden_renormalization_2019}. (i) Let $P^{\mu \nu}_{ij}$ with $1 \leq i,j \leq L$ be a plane of plaquettes extending in the directions $\mu$ and $\nu$. The operators
\begin{equation}
    \mathcal{P}^{\mu \nu}_{ij} \equiv  \prod_{p \in P^{\mu \nu}_{ij}} \sigma^z_p
\end{equation}
commute with all $A_{\ell}$ and all $B_c$ and act nontrivially on the ground state $\ket{\psi_0}_{\tc}$. (ii) Similarly, let $Q^{\mu \nu}_{ij}$ for $1 \leq i,j \leq L$ be the set of all $\mu-\nu$ plaquettes with fixed $\mu$ coordinate $i$ and $\nu$ coordinate $j$. The  operators 
\begin{equation}
\mathcal{Q}^{\mu \nu}_{ij}  \equiv \prod_{p \in Q^{\mu \nu}_{ij}} \sigma^x_p
\end{equation}
commute with all $A_{\ell}$ and $B_c$ and act nontrivially on the ground state $\ket{\phi_0}$. Note that $\mathcal{P}^{\mu \nu}_{ij}$ and $\mathcal{Q}^{\rho \sigma}_{k \ell}$ anticommute for $(\mu , \nu) = (\rho, \sigma)$ and commute otherwise. This allows us to generate the entire code space of dimension 64 from either ground state (\ref{eq:psiphidefn}), in a manner completely analogous to that of the 2DTC's $d=1$ gauge-like symmetries \cite{kitaev_fault-tolerant_2003, nussinov_autocorrelations_2008}.

\subsection{Haah's Code}
Notably, Haah's Code does not have any string-like logical operators, and therefore has no $d=1$ dimensional gauge-like symmetries. Instead, it has the following two $d=2$ dimensional gauge-like symmetries \cite{haah_local_2011}: (i) let $P^{\mu}_i$ for $1 \leq i \leq L$ be a plane of vertices perpendicular to the $\mu$ direction. Then, 
\begin{equation}\mathcal{P}^{\mu}_i \equiv \prod_{v \in P^{\mu}_i} \sigma^z_v
\end{equation}
commute with each $A_v$ and $B_v$ and are symmetries. These operators act nontrivially on $\ket{\psi_0}$. Similarly, the symmetries 
\begin{equation}
    \mathcal{Q}^{\mu}_i \equiv \prod_{v \in P^{\mu}_i} \tau^z_v
\end{equation}
commute with $A_v$ and $B_v$ and alter the ground state $\ket{\phi_0}$. Unlike the 4DTC, these logical operators commute, and do not constitute all logical operators of the code \cite{haah_local_2011}.

\section{4D Toric Code Duality}
In this section, we more carefully prove the duality given by equation (12) of the main paper. We start by defining the model as in the main paper: let $\Lambda = \mathbb{Z}_L^4$ be a $d=4$ lattice of length $L$ in each direction, with sites $v = (v_1 , v_2, v_3, v_4) \in \Lambda$. This lattice has $L^4$ vertices $v$, $4L^4$ 1D links $\ell$, $6L^4$ 2D plaquettes $p$, $4L^4$ 3D cubes $c$, and $L^4$ 4D hypercubes $h$. We associate with each plaquette a two-dimensional Hilbert space $\mathcal{H}_p = \mathbb{C}^2$, and with the full lattice a total Hilbert space $\mathcal{H} = \otimes_p \mathcal{H}_p$. We use the operators $\sigma^{\mu}_p$ with $\mu = x,y,z$ to denote the $\mu$-Pauli operator acting on $\mathcal{H}_p$. Finally, we define $A_{\ell}$, $B_c$, $H_{\tc}$, $\Z_{\tc}$, $\T_a$, and $\T_b$ as in equations (2) through (8) of the main paper.

We start by noting that the $A_{\ell}$ and $B_c$ algebras are identical. This is because the center of every plaquette in the lattice also lies at the center of every plaquette of the dual lattice, as can easily be verified. In the dual lattice, $A_{\ell}$ looks like the product of plaquettes in a cube, while $B_c$ looks like the product of plaquettes surrounding a link (c.f. the 2D toric code \cite{kitaev_fault-tolerant_2003}, for which the operators $A_s$ and $B_p$ are interchanged in roles on the dual lattice). As a result, $\T_a$ and $\T_b$ will be identical, with $a \leftrightarrow b$. We will focus on the $B_c$ algebra, which may be easier to visualize (despite its four-dimensionality).

The form of $B_c$ constraints are most easily described in the language of homology. For a brief pedagogical introduction to homology and their relation to toric codes, see Appendix A of Ref. \cite{anwar_fast_2014}. In short, each $B_c$ is associated with an oriented 3-cell (cube) $c$ on the 4-torus, and $B_c$ products are associated with 3-chains, formal sums of 3-cells over $\mathbb{Z}_2$:
\begin{equation}
    \prod_c B_c^{n_c} \Longleftrightarrow \sum_c n_c c , \quad n_c = 0,1 .
\end{equation}
To obtain such a product in terms of $\sigma^z_p$ operators, we apply the boundary operator $\partial_3$, which yields a formal sum of 2-cells (plaquettes) $p$:
\begin{equation}
    \partial_3 \qty[ \sum_c n_c c ] = \sum_c n_c \partial_3 c = \sum_c n_c \sum_{p \in \partial c} p ,
\end{equation}
A set of 3-cells $P$ is therefore a constraint exactly when its corresponding 3-chain is a 3-cycle -- that is, a 3-chain with no boundary:
\begin{equation}
    \prod_{c \in P} B_c = \id \Longleftrightarrow \partial_3 \sum_{c \in P} c = 0
\end{equation}
Let $\Gamma$ be the set of all cellular 3-cycles on $\Lambda$. Then, $\T_b$ is given by:
\begin{equation}
    \T_b = \sum_{\gamma \in \Gamma} \Tb^{|\gamma|} = \sum_{n=0}^{4L^4} b_n \Tb^n ,
\end{equation}
where $|\gamma|$ is the hypersurface area of the 3-cycle $\gamma$, and each $b_n$ is the number of cellular 3-cycles of hypersurface area $n$ (including the empty 3-cycle, so that $b_0 = 1$). The simplest such 3-cycle is $\partial h$, the boundary of a simple hypercube $h$. One can easily verify that the product of $B_c$ over the eight cubes in the hypercube yields the identity, as given by equation (13) of the main paper.

On the other hand, consider the 4D Ising model given by equations (9) and (10) of the main paper. For any spin configuration $\C$, we may draw a \textit{domain wall} in the dual lattice around each connected set of flipped spins. These domain walls are also cellular 3-cycles on the lattice, with an extra condition: domain walls are necessarily the boundary of a 4-chain, the hypervolume of spins flipped. As a result, domain walls are required to be topologically trivial, or \textit{contractible}, 3-cycles. Letting $\Gamma^*$ be the set of contractible 3-cycles, $\Z_{\is}$ can be written:
\begin{equation}
\label{eq:IsingExpansion}
\begin{split}
    \Z_{\is} &= 2e^{4L^4 \beta J} \sum_{\gamma^* \in \Gamma^*} \qty(e^{-2\beta J})^{|\gamma^*|} \\
    &= 2e^{4L^4 \beta J} \sum_{n=0}^{4L^4} b_n^* \qty(e^{-2\beta J})^n ,
\end{split}
\end{equation}
where $b_n^*$ is the number of contractible 3-cycles of hypersurface area $n$. We note that all contractible 3-cycles can be formed from the boundaries of simple hypercubes -- that is, any domain wall can be formed by combining those from individual spin flips.

By contrast, $\T_b$ also contains topologically nontrivial 3-cycles, stretching across the periodic boundaries. However, these 3-cycles necessarily contain no less than $L^3$ 3-cells, in order to stretch across the torus and eliminate the 2-cell boundaries. We therefore find that $b_n = b_n^*$ for $n < L^3$, proving equation (12) of the main paper.

It is now necessary to show that the $\mathcal{O}(\Tb^{L^3})$ terms are indeed negligible in the thermodynamic limit. First, we remark that these terms vanish under different boundary conditions: had our lattice been endowed with three or fewer periodic boundaries, then all 3-cycles would be contractible, and the duality would be exact for all system sizes. Since the thermodynamic free energy should not depend on boundary conditions, we expect on physical grounds that the $\mathcal{O}(\Tb^{L^3})$ terms should not contribute in the thermodynamic limit. Indeed, since $\Tb^{L^3} \rightarrow 0$ very quickly for all $\beta$, any coefficients $b_n$ must scale quite pathologically with $L$ to have any hope of contributing in the thermodynamic limit.

To prove that the topologically nontrivial terms of $\T_b$ are trivial in the thermodynamic limit, we use the following standard observation from homology: there exist only four homologically distinct nontrivial 3-cycles on the 4-torus. Two nontrivial 3-cycles are considered homologically equivalent if they differ by the boundary of a 4-chain -- that is, a contractible 3-cycle. As a result, every possible 3-cycle $\Sigma$ can be written as:
\begin{equation}
    \Sigma = \sum_c n_c c + \sum_{m=1}^4 C_m
\end{equation}
where each $C_m$ is one of four representative nontrivial 3-cycles from each homology class. A more careful calculation of $\T_b$ then gives:
\begin{equation}
    \T_b = \sum_{\gamma^* \in \Gamma^*} \qty{ \Tb^{|\gamma^*|} \qty[1+ \sum_{m=1}^4 \Tb^{L^3 - |\gamma^* \cap C_m|} + \ldots  ] } ,
\end{equation}
where the ellipsis contains the additional 11 terms corresponding to two, three, or all four $C_m$. While it is difficult in general to determine the order of terms with nontrivial cycles, we can bound $\T_b$ above by replacing a given order of $\Tb$ with \textit{all} orders of $\Tb$, giving a summable expression:
\begin{equation}
    \begin{split}
    \T_b &\leq \sum_{\gamma^* \in \Gamma^*} \qty{ \Tb^{|\gamma|} \qty[1+15\sum_{n=0}^{4L^4} \Tb^n ] } \\
    &= \qty[ 1+15\sum_{n=0}^{4L^4} \Tb^n ] \T^*_b ,
    \end{split}
\end{equation}
where $\T_b^* = \sum b_n^* \Tb^n$. In the free energy, we therefore have:
\begin{equation}
    \frac{1}{L^4} \log \T_b \leq \frac{1}{L^4} \log \T_b^* + \frac{1}{L^4} \log \qty[\frac{16 - \Tb}{1-\Tb}]
\end{equation}
Since $\T_b$ is also bounded below by $\T_b^*$, we have proven that the two give the same thermodynamic free energy. This gives the desired 4DTC-Ising duality.

As a final remark, although the $\mathcal{O}(\Tb^{L^3})$ terms above appear due to the topology of the lattice, they do not contain any of the information regarding the topological degeneracy associated with the 4DTC. For example, the 2DTC on a simple torus of genus $g = 1$ is dual to two Ising chains \cite{nussinov_autocorrelations_2008}. Since each Ising chain is two-fold degenerate, the conventional 2DTC on a torus is four-fold degenerate. Further analysis employing the algebraic relations between the ``bonds" \cite{nussinov_symmetry_2009, nussinov_effective_2012, nussinov_sufficient_2009, cobanera_bond-algebraic_2011, weinstein_absence_2018, PhysRevB.86.085415} and the Euler-L'Huillier relation reveals that for the 2DTC, on general $g>0$ manifolds, the degeneracy of all levels (not only of those in the ground state sector) is $4^{g-1}$ times that of two decoupled Ising chains \cite{nussinov_symmetry_2009}. This constant then correspondingly appears as a global prefactor in the dual partition function. We similarly expect that for the 4DTC, the topological degeneracy commonly associated with the ground state manifold indeed extends to all energy levels of the system.

\section{A duality mapping preserving the locality of heat bath couplings}
A common criticism of using dualities to investigate dynamics is that local perturbations become generally nonlocal in the dual model. As a result, a coupling to a local heat bath modeled by a local magnetic field might in general become a highly nonlocal coupling in the dual model, and it may not be possible to infer dynamics of the original model from the dual model as a result. Here, we present a proof-of-concept dualization of a local heat bath to a local heat bath, in order to show that it is possible to construct dualities which should certainly preserve local dynamics. We do not show that \textit{all} local heat baths can be mapped to a dual local heat bath -- indeed, local heat baths will generically become highly nonlocal. Instead, by showing that these mappings are possible, we argue that the original and dual models in most dualities should have sufficiently similar thermalization on physical grounds.

We consider the ordinary 2D Toric Code \cite{kitaev_fault-tolerant_2003} on a square lattice of side length $L$, with stabilizer generators given by:
\begin{equation}
\begin{split}
A_s = \prod_{\ell \in s} \sigma^x_{\ell} , \\
B_p = \prod_{\ell \in p} \sigma^z_{\ell} ,
\end{split}
\end{equation}
where each $s$ is a ``star" of four links surrounding a given vertex, and each $p$ is a ``plaquette" of four links forming a square. The unperturbed stabilizer Hamiltonian is given by:
\begin{equation}
\label{H0}
H_0 = -a \sum_s A_s - b \sum_p B_p .
\end{equation}

In the following, we will employ open (i.e., non-periodic) boundary conditions, in which the stars on the boundary will not be included in the sum (\ref{H0}). We note that the Toric Code is traditionally defined on a torus, but that the thermodynamics  in the bulk of the system should not be affected by the boundary conditions in the thermodynamic limit.

For notational convenience, we will from now on relabel operators by their $(i,j)$ coordinates: 
\begin{equation}
\begin{split}
A_s = A_{i,j} , \quad 1 \leq i,j \leq L-1 , \\
B_p = B_{i,j} , \quad 0 \leq i,j \leq L-1 ,
\end{split}
\end{equation}
where the $(i,j)$ coordinates of $s$ denote the lattice site of the star, and the $(i,j)$ coordinates of $p$ denote the bottom-left corner of $p$. Finally, we will denote the coordinates of spins on links as if they were placed in the center of their respective link. That is, Pauli operators on $x$-links will be denoted by:
\begin{equation}
\sigma^{\mu}_{i+\frac{1}{2} , j}, \quad 0 \leq i \leq L-1, \quad 0 \leq j \leq L,
\end{equation}
and Pauli operators on $y$-links will be denoted by:
\begin{equation}
\sigma^{\mu}_{i, j+\frac{1}{2}}, \quad 0 \leq i \leq L, \quad 0 \leq j \leq L-1.
\end{equation}

As a first simple example, we consider a local bath modeled by a perturbing magnetic field acting on only $x$-links, given by:
\begin{equation}
\label{Vx}
V_x = -\lambda \sum_{i=0}^{L-1} \sum_{j=0}^{L} \sigma^x_{i+\frac{1}{2} , j} ,
\end{equation}
where $\lambda > 0$ gives the magnetic field strength. Each $\sigma^x$ in (\ref{Vx}) anticommutes with two adjacent plaquettes, and commutes with all stars. The commutation relations of the full Hamiltonian $H_0 + V_x$ is then given by:
\begin{equation}
\label{algebra}
\begin{split}
\comm{A_s}{A_{s'}} = \comm{B_p}{B_{p'}} = \comm{A_s}{B_p} = \comm{A_s}{\sigma^x_{i+\frac{1}{2},j}} = 0 , \\
\acomm{B_{i,j}}{\sigma^x_{i+\frac{1}{2},j}} = \acomm{B_{i,j}}{\sigma^x_{i+\frac{1}{2},j+1}} = 0 .
\end{split}
\end{equation}  We note that a magnetic field acting only on $y$-links, or a field coupling to $\sigma^z$ instead of $\sigma^x$, would behave in very much the same manner. 

To employ a bond-algebraic duality, we map each operator in (\ref{H0}) and (\ref{Vx}) to another set of spin operators $\tau^{\mu}$ in such a way as to preserve the bond algebra (\ref{algebra}). Towards this end, we map each vertical column of plaquettes to an open Ising chain, and map (\ref{Vx}) to a transverse field on these chains:
\begin{equation}
\begin{split}
B_{i,j} &\rightarrow \tau^z_{i,j} \tau^z_{i,j+1} , \\
\sigma^x_{i+\frac{1}{2},j} &\rightarrow \tau^x_{i,j} .
\end{split}
\end{equation}
The stars can then each be mapped to auxiliary spins $\tau^z_{L,j}$. The resulting dual Hamiltonian is given by:
\begin{equation}
\begin{split}
&H_0 + V_x \rightarrow H_x^{\text{Dual}}\\
&= \sum_{i=0}^{L-1} \left[ \sum_{j=0}^{L-1} \tau^z_{i,j} \tau^z_{i,j+1} + \sum_{j=0}^L \tau^x_{i,j} \right] + \sum_{j=1}^{(L-1)^2} \tau^z_{L,j} .
\end{split}
\end{equation}

We might also consider a local bath modeled by a perturbing field that anticommutes with both stars and plaquettes, given by:
\begin{equation}
V_y = - \lambda \sum_{i=0}^{L-1} \sum_{j=0}^L \sigma^y_{i+\frac{1}{2}, j} .
\end{equation}
Each $\sigma^y_{i+\frac{1}{2},j}$ once again anticommutes with vertically adjacent plaquettes, but now also anticommutes with horizontally adjacent stars. The commutation relations are given by:
\begin{equation}
\begin{split}
\comm{A_s}{A_{s'}} = \comm{B_p}{B_{p'}} = \comm{A_s}{B_p} &= 0 , \\
\acomm{B_{i,j}}{\sigma^y_{i+\frac{1}{2},j}} = \acomm{B_{i,j}}{\sigma^y_{i+\frac{1}{2},j+1}} &= 0, \\
\acomm{A_{i,j}}{\sigma^y_{i+\frac{1}{2},j}} = \acomm{A_{i,j}}{\sigma^y_{i-\frac{1}{2},j}} &= 0 .
\end{split}
\end{equation}
Here, each vertical column of plaquettes is once again mapped to an open Ising chain of $\tau^z$ spins, while horizontal rows of stars are mapped to open Ising chains of $\rho^z$ spins. To anticommute with both $\tau$ chains and $\rho$ chains, each $\sigma^y$ is mapped to a transverse coupling between chains $\tau^x \rho^x$:
\begin{equation}
\begin{split}
B_{i,j} \rightarrow \tau^z_{i,j} \tau^z_{i,j+1} , \\
A_{i,j} \rightarrow \rho^z_{i,j} \rho^z_{i+1,j} , \\
\sigma^y_{i+\frac{1}{2},j} \rightarrow \tau^x_{i,j} \rho^x_{i,j} .
\end{split}
\end{equation}
%note: consider A's on boundary in mapping. Potentially not quite right
The resulting dual Hamiltonian is then:
\begin{equation}
\begin{split}
&H_0 + V_y \rightarrow H_y^{\text{Dual}} \\
&= \sum_{i=0}^{L-1} \left[ \sum_{j=0}^{L-1} \tau^z_{i,j} \tau^z_{i,j+1} \right] + \sum_{j=1}^{L-1} \left[ \sum_{i=1}^{L-1} \rho^z_{i,j} \rho^z_{i+1,j} \right] \\
&+ \sum_{i=0}^{L-1} \sum_{j=0}^L \tau^x_{i,j} \rho^x_{i,j} .
\end{split}
\end{equation}
At first glance, the above duality might seem highly nonlocal due to the coupling terms $\tau^x \rho^x$. However, if the $\tau$ Ising chains are visualized as vertical Ising chains, and the $\rho$ Ising chains are visualized as horizontal Ising chains lying across the $\tau$ Ising chains in a crossing pattern, then the couplings are localized to the points where the Ising chains cross.

\bibliography{refs}